\documentstyle[12pt,a4]{article}
\newcommand\lsim{\mathrel{\rlap{\lower4pt\hbox{\hskip1pt$\sim$}}
    \raise1pt\hbox{$<$}}}
\newcommand\gsim{\mathrel{\rlap{\lower4pt\hbox{\hskip1pt$\sim$}}
    \raise1pt\hbox{$>$}}}

\catcode`\@=11
\def\marginnote#1{}
\def\ifmath#1{\relax\ifmmode #1\else $#1$\fi}

\def\stop{\,\widetilde{t}}

\def\bold#1{\setbox0=\hbox{$#1$}%
     \kern-.025em\copy0\kern-\wd0
     \kern.05em\copy0\kern-\wd0
     \kern-.025em\raise.0433em\box0 }

\def\GENITEM#1;#2{\par\vskip6pt \hangafter=0 \hangindent=#1
   \Textindent{$ #2$ }\ignorespaces}

\newcount\hour
\newcount\minute
\newtoks\amorpm
\hour=\time\divide\hour by60
\minute=\time{\multiply\hour by60 \global\advance\minute by-
\hour}
\edef\standardtime{{\ifnum\hour<12 \global\amorpm={am}%
    \else\global\amorpm={pm}\advance\hour by-12 \fi
    \ifnum\hour=0 \hour=12 \fi
    \number\hour:\ifnum\minute<100\fi\number\minute\the\amorpm}}
\edef\militarytime{\number\hour:\ifnum\minute<100\fi\number\minute}
\def\draftlabel#1{{\@bsphack\if@filesw {\let\thepage\relax
  \xdef\@gtempa{\write\@auxout{\string
    \newlabel{#1}{{\@currentlabel}{\thepage}}}}}\@gtempa
    \if@nobreak \ifvmode\nobreak\fi\fi\fi\@esphack}
     \gdef\@eqnlabel{#1}}
\def\@eqnlabel{}
\def\@vacuum{}
\def\draftmarginnote#1{\marginpar{\raggedright\scriptsize\tt#1}}
\def\draft{\oddsidemargin -.5truein
        \def\@oddfoot{\sl preliminary draft \hfil
        \rm\thepage\hfil\sl\today\quad\militarytime}
        \let\@evenfoot\@oddfoot \overfullrule 3pt
        \let\label=\draftlabel
        \let\marginnote=\draftmarginnote

\def\@eqnnum{(\theequation)\rlap{\kern\marginparsep\tt\@eqnlabel}%
\global\let\@eqnlabel\@vacuum}  }
\def\preprint{\twocolumn\sloppy\flushbottom\parindent 1em
        \leftmargini 2em\leftmarginv .5em\leftmarginvi .5em
        \oddsidemargin -.5in    \evensidemargin -.5in
        \columnsep 15mm \footheight 0pt
        \textwidth 250mmin      \topmargin  -.4in
        \headheight 12pt \topskip .4in
        \textheight 175mm
        \footskip 0pt

\def\@oddhead{\thepage\hfil\addtocounter{page}{1}\thepage}
        \let\@evenhead\@oddhead \def\@oddfoot{} \def\@evenfoot{}
}
\def\titlepage{\@restonecolfalse\if@twocolumn\@restonecoltrue\o
necolumn
     \else \newpage \fi \thispagestyle{empty}\c@page\z@
        \def\thefootnote{\fnsymbol{footnote}} }
\def\endtitlepage{\if@restonecol\twocolumn \else  \fi
        \def\thefootnote{\arabic{footnote}}
        \setcounter{footnote}{0}}  %\c@footnote\z@ }
\catcode`@=12
\relax
\def\be{\begin{equation}}
\def\ee{\end{equation}}
\def\bea{\begin{eqnarray}}
\def\eea{\end{eqnarray}}

\def\NPB#1#2#3{{\it Nucl.~Phys.} {\bf{B#1}} (19#2) #3}
\def\PLB#1#2#3{{\it Phys.~Lett.} {\bf{B#1}} (19#2) #3}
\def\PRD#1#2#3{{\it Phys.~Rev.} {\bf{D#1}} (19#2) #3}
\def\PRL#1#2#3{{\it Phys.~Rev.~Lett.} {\bf{#1}} (19#2) #3}
\def\ZPC#1#2#3{{\it Z.~Phys.} {\bf C#1} (19#2) #3}

\def\AP#1#2#3{{\it Ann.~Phys.} {\bf#1} (19#2) #3}
\def\RMP#1#2#3{{\it Rev.~Mod.~Phys.} {\bf#1} (19#2) #3}

\def\mst11{m_{\;\widetilde{t}_{1}}}

\def\mst22{m_{\;\widetilde{t}_{2}}}
\def\mst12{m_{\;\widetilde{t}_{1,2}}}

\def\msb11{m_{\;\widetilde{b}_{1}}}
\def\msb22{m_{\;\widetilde{b}_{2}}}
\def\msb12{m_{\;\widetilde{b}_{1,2}}}

\def\mwidetilde2{\widetilde{m}^{2}}

\relax
\begin{document}
\topmargin-2.5cm
%\draft
%\preprint
%
\begin{titlepage}
\begin{flushright}
CERN-TH/98-81\\
OUTP-98-23-P\\
\end{flushright}
\vskip 0.3in
\begin{center}
{\Large\bf  The More Relaxed}
\vskip 0.1in
{\Large\bf Supersymmetric Electroweak Baryogenesis}
\vskip .5in

{\bf A. Riotto\footnote{On leave of absence from Department of Physics, Theoretical  Physics, University of Oxford, U.K. Email: riotto@nxth04.cern.ch.}    }
\vskip.35in
 CERN, TH Division\\
CH-1211 Geneva 23, Switzerland 
\end{center}
\vskip 0.5cm
\begin{center}
{\bf Abstract}
\end{center}
\begin{quote}

We reanalyze the issue of generation of  the baryon asymmetry at  the electroweak phase transition in  the  MSSM and  compute the baryon asymmetry assuming the presence of non-trivial CP-violating phases in the parameters associated with the left-right stop mixing term and the Higgsino mass $\mu$.
Making use of the closed time-path (CTP)  formalism of nonequilibrium field theory,  we write down a set of quantum Boltzmann equations describing the local number density asymmetries of the particles involved in supersymmetric electroweak baryogenesis. CP-violating sources manifest ``memory'' effects which are  typical of the quantum transport theory and are not present in the classical approach. Compared to previous estimates, these non-Markovian features enhance the final baryon asymmetry by at least   two orders of magnitude. This means that  a  CP-violating phase $|\sin \phi_{\mu}|$ as small as $10^{-3}$ (or even smaller) is enough to generate the observed baryon asymmetry.

\end{quote}
\vskip 0.5cm
\begin{flushleft}
%CERN-TH/96-242\\
March 1998 \\
\end{flushleft}

\end{titlepage}
\setcounter{footnote}{0}
\setcounter{page}{0}
\newpage
\baselineskip=20pt

\begin{flushleft}
{\bf  1. Introduction and summary}
\end{flushleft}

The presence of  unsuppressed
baryon number violating processes at high temperatures within  
the Standard Model (SM) of weak interactions makes the  
generation of the baryon number at the electroweak scale an appealing scenario \cite{reviews}. The baryon number violating processes    also impose severe constraints on  
models where the baryon asymmetry is created at energy scales much higher than the electroweak scale \cite{anomaly}. Unfortunately, the electroweak phase transition  is too weak  in the SM \cite{transition}. This   means that the baryon asymmetry  
generated during the transition would subsequently be  erased by unsuppressed  
sphaleron transitions in the broken phase. 
 The most promising and  
well-motivated framework for electroweak baryogenesis beyond the SM  seems to be supersymmetry (SUSY).  Electroweak  
baryogenesis in the framework of the Minimal Supersymmetric Standard Model  
(MSSM) has  attracted much attention in the past years, with 
particular emphasis on the strength of the phase transition ~\cite{early1,early2,early3} and  
the mechanism of baryon number generation \cite{nelson,noi,higgs,ck}.

Recent  analytical \cite{r1,r2} and  lattice  
computations  \cite{r3} have  revealed   that the phase transition can be sufficiently strongly  
first order if  the
ratio of the vacuum expectation values of the two neutral Higgses $\tan\beta$  
is smaller than $\sim 4$. Moreover, taking into account all the experimental bounds 
as well as those coming from the requirement of avoiding dangerous
 color breaking minima,  the lightest Higgs boson should be  lighter than about  $105$ GeV,
 while the right-handed stop mass might  be close to the present experimental bound and should
 be smaller than, or of the order of, the top quark mass \cite{r2}. 

Moreover, the MSSM contains additional sources  
of CP-violation  besides the CKM matrix phase. 
These new phases are essential  for the generation of the baryon number since  large  
CP-violating sources may be  locally induced by the passage of the bubble wall separating the broken from the unbroken phase during the electroweak phase transition.   Baryogenesis is fuelled  when transport properties allow the CP-violating  
charges to efficiently diffuse in front of the advancing bubble wall where
anomalous electroweak baryon violating processes are not suppressed.
The new phases   
appear    in the soft supersymmetry breaking  
parameters associated to the stop mixing angle and to  the gaugino and  
neutralino mass matrices; large values of the stop mixing angle
are, however, strongly restricted in order to preserve a
sufficiently strong first order electroweak phase transition. 
Therefore, an acceptable baryon asymmetry from the stop sector
may only be generated through a delicate balance between the values
of the different soft supersymmetry breaking parameters contributing
to the stop mixing parameter, and their associated CP-violating 
phases \cite{noi}. As a result, the contribution to the final baryon asymmetry from the stop sector turns out to be negligible.   On the other hand, charginos and neutralinos may be responsible for the observed baryon asymmetry  \cite{noi,ck}. Yet, 
this is true within the MSSM. If the strength of the
 electroweak phase transition is enhanced by the presence of some
new degrees of freedom beyond the ones contained in the MSSM, {\it
e.g.} some extra standard model gauge singlets,
 light stops (predominantly the 
right-handed ones) and charginos/neutralinos are expected to 
give quantitatively the
same contribution to the final baryon asymmetry.

While in the past few years a lot of effort has been devoted  to the study of the strength of the phase transition within the MSSM, only very recently some attention has been paid to the mechanism by which the baryon asymmetry is produced. 
The baryon asymmetry has been usually computed using a number of steps  \cite{nelson,noi,riotto}: 
{\it 1)}  look for 
 those  charges which are
approximately conserved in the symmetric phase, so that they
can efficiently diffuse in front of the bubble where baryon number
violation is fast, and non-orthogonal to baryon number,
so that the generation of a non-zero baryon charge is energetically
favoured.
Charges with these characteristics in the MSSM 
are the axial stop charge and the Higgsino charge,
which may be produced from the interactions of squarks and
charginos
and/or neutralinos with the bubble wall,
provided a source of CP-violation is present in these sectors; 
{\it 2)}  compute the CP-violating currents  of the plasma locally induced by the passage of the bubble wall; 
{\it 3)} write and solve a set of coupled differential diffusion equations for the local particle densities, including the CP-violating source terms derived from the computation of the current at  step {\it 2)}  and the particle number changing reactions. The solution to these equations gives a net baryon number which is produced in the symmetric phase and then transmitted into the interior of the bubbles of the broken phase, where it is not wiped out if the first transition is strong enough. 
Notice that the CP-violating sources are  inserted into the diffusion equations by hand only after the CP-violating currents have been defined and computed. 
This procedure introduces    some degree of  arbitrariness and --indeed--  different  CP-violating sources have been adopted  for the stop and the Higgsino sectors in the literature \cite{nelson,noi}. This is not an academic question.  Adopting different sources  leads to   different numerical results for the final baryon asymmetry, especially if  the sources are  expressed in terms of   a different number of derivatives of the Higgs bubble wall profile and, therefore, in terms of  different powers of the bubble wall velocity $v_\omega$ and bubble wall width $L_\omega$. 

This crucial issue has been recently investigated  in \cite{riotto97} where it was shown that  non-equilibrium Quantum Field Theory  provides us with   the necessary    tools  to  write down a set of quantum Boltzmann equations (QBE's) describing the local particle densities and automatically incorporating the CP-violating sources. The most appropriate extension of the field theory
to deal with these issues is to generalize the time contour of
integration to a closed time-path (CTP).  The CTP formalism is a powerful Green's function
formulation for describing non-equilibrium phenomena in field theory, it leads to 
 a complete
non-equilibrium quantum kinetic theory approach and   to   a   rigorous   computation of  the CP-violating sources for the stop and the Higgsino numbers.   
In this way, the latter  have been  rigorously defined and the   level  of  arbitrariness  of the  previous treatments  has been  dismissed \cite{riotto97}.  What is more relevant, though, is that 
the CP-violating sources-- and more generally the particle number changing interactions--  built up from the CTP formalism are characterized by  ``memory'' effects which are typical of  the  quantum transport theory \cite{dan,henning}. CP-violating sources are built up when right-handed stops and Higgsinos scatter off the advancing Higgs bubble wall and CP is violated at the vertices of interactions.
In  the classical kinetic theory the ``scattering term'' does not include any integral over the past history of the system. This   is equivalent to assuming that any collision in the plasma  does not depend upon the previous ones.   On the contrary, the quantum approach reveals that the CP-violating source is manifestly non-Markovian. As we shall see, these memory effects enhance the value of of the final baryon asymmetry  by at least two orders of magnitude with respect to the previous results. This means that the lower bound on the CP-violating phases from requiring sueccessful baryogenesis is considerably relaxed --phases as large as  $10^{-3}$ or smaller are enough to generate
the observed baryon symmetry.   This has  important implications for the supersymmetric CP-problem.
We will also investigate  the structure of the   kinetic  QBE's derived with the CTP formalism. These equations have an obvious interpretation in terms of gain and loss processes.   However,   the equations are manifestly non-Markovian and only the assumption that the relaxation time scale of the particle asymmetry is much longer than the time scale of the non-local kernels leads to a Markovian description.  Further approximations lead to the familiar Boltzmann  equations.

The paper is organized as follows. In  section 2  we give a brief description of the basic features of the non-equilibrium quantum field theory and the CTP formalism. In sections 3  and 4  we compute the quantum transport equations for local particle asymmetries in the bosonic and fermion case, respectively. Section 5   contains the explicit 
computation of the CP-violating source for the right-handed stop  and the
discussion about how to go from general QBE's to diffusion/Boltzmann equations. Section 6 is devoted to the same issues, but for the Higgsino number. Section 7 is devoted to the computation of the final baryon asymmetry and comparison to previous results. 
 We conclude with an outlook of our findings in section 8.

\begin{flushleft}
{\bf  2. The Schwinger-Keldysh formalism for  non-equilibrium quantum field theory}
\end{flushleft}

In this section  we will   briefly present    some of the  basic features of the  non-equilibrium quantum field theory based on the Schwinger-Keldysh formulation \cite{sk}. The interested reader is referred to the excellent review by Chou   {\it et al.} \cite{chou} for a more exhaustive discussion.

Since  we
need the temporal evolution of the particle asymmetries with definite initial conditions and not
simply the transition amplitude of particle reactions, 
the ordinary equilibrium quantum field theory at finite temperature   is not the appropriate tool. 
The most appropriate extension of the field theory
to deal with nonequilibrium phenomena amounts to generalize the time contour of
integration to a closed-time path. More precisely, the time integration
contour is deformed to run from $-\infty$ to $+\infty$ and back to
$-\infty$.  

 The CTP formalism (often  dubbed as in-in formalism) is a powerful Green's function
formulation for describing non-equilibrium phenomena in field theory.  It
allows to describe phase-transition phenomena and to obtain a
self-consistent set of quantum Boltzmann equations.
The formalism yields various quantum averages of
operators evaluated in the in-state without specifying the out-state. On the contrary, the ordinary quantum field theory (often dubbed as in-out formalism) yields quantum averages of the operators evaluated  with an in-state at one end and an out-state at the other. 

Because of the time contour deformation, the partition function in the in-in formalism for a {\it complex} scalar field is defined to be
\begin{eqnarray}
Z\left[ J, J^{\dagger}\right] &=& {\rm Tr}\:\left[ T\left( {\rm exp}\left[i\:\int_C\:\left(J\phi+J^{\dagger}\phi^{\dagger} \right)\right]\right)\rho\right]\nonumber\\
&=& {\rm Tr}\:\left[ T_{+}\left( {\rm exp}\left[ i\:\int\:\left(J_{+}\phi_{+}+J^{\dagger}_{+}\phi^{\dagger}_{+} \right)\right]\right)\right.
\nonumber\\
&\times&\left.  T_{-}\left( {\rm exp}\left[ -i\:\int\:\left(J_{-}\phi_{-}+J^{\dagger}_{-}\phi^{\dagger}_{-} \right)\right]\right) \rho\right],
\end{eqnarray}
where the suffic $C$ in the integral denotes that the time integration contour runs from minus infinity to plus infinity and then back to minus infinity again. The symbol $\rho$ represents the initial density matrix and the fields are in the Heisenberg picture  and  defined on this closed time contour. 
As with the Euclidean time formulation, scalar (fermionic) fields $\phi$ are
still periodic (anti-periodic) in time, but with
$\phi(t,\vec{x})=\phi(t-i\beta,\vec{x})$, $\beta=1/T$.
The temperature appears   due to boundary
condition, but time is now  explicitly present in the integration
contour.

We must now identify field
variables with arguments on the positive or negative directional
branches of the time path. This doubling of field variables leads to
six  different real-time propagators on the contour \cite{chou}.  These six
propagators are not independent, but using all of them simplifies the notation. 
For a generic bosonic charged  scalar field $\phi$ they are defined as 
\begin{eqnarray}
\label{def1}
G_{\phi}^{>}\left(x, y\right)&=&-i\langle
\phi(x)\phi^\dagger (y)\rangle,\nonumber\\
G_{\phi}^{<}\left(x,y\right)&=&-i\langle
\phi^\dagger (y)\phi(x)\rangle,\nonumber\\
G^t _{\phi}(x,y)&=& \theta(x,y) G_{\phi}^{>}(x,y)+\theta(y,x) G_{\phi}^{<}(x,y),\nonumber\\
G^{\bar{t}}_{\phi} (x,y)&=& \theta(y,x) G_{\phi}^{>}(x,y)+\theta(x,y) G_{\phi}^{<}(x,y), \nonumber\\
G_{\phi}^r(x,y)&=&G_{\phi}^t-G_{\phi}^{<}=G_{\phi}^{>}-G^{\bar{t}}_{\phi}, \:\:\:\: G_{\phi}^a(x,y)=G^t_{\phi}-G^{>}_{\phi}=G_{\phi}^{<}-G^{\bar{t}}_{\phi},
\end{eqnarray}
where the last two Green functions are the retarded and advanced Green functions respectively and $\theta(x,y)=\theta(t_x-t_y)$ is the step function.  For a generic fermion field $\psi$ the six different propagators are analogously defined as
\begin{eqnarray}
\label{def2}
G^{>}_{\psi}\left(x, y\right)&=&-i\langle
\psi(x)\bar{\psi} (y)\rangle,\nonumber\\
G^{<}_{\psi}\left(x,y\right)&=&+i\langle
\bar{\psi}(y)\psi(x)\rangle,\nonumber\\
G^{t}_{\psi} (x,y)&=& \theta(x,y) G^{>}_{\psi}(x,y)+\theta(y,x) G^{<}_{\psi}(x,y),\nonumber\\
G^{\bar{t}}_{\psi} (x,y)&=& \theta(y,x) G^{>}_{\psi}(x,y)+\theta(x,y) G^{<}_{\psi}(x,y),\nonumber\\
G^r_{\psi}(x,y)&=&G^{t}_{\psi}-G^{<}_{\psi}=G^{>}_{\psi}-G^{\bar{t}}_{\psi}, \:\:\:\: G^a_{\psi}(x,y)=G^{t}_{\psi}-G^{>}_{\psi}=G^{<}_{\psi}-G^{\bar{t}}_{\psi}.
\end{eqnarray}
For equilibrium phenomena, the brackets $\langle \cdots\rangle$ imply a thermodynamic average over all the possible states of the system. While for homogeneous systems in equilibrium, the Green functions
depend only upon the difference of their arguments $(x,y)=(x-y)$ and there is no dependence upon $(x+y)$,  for systems out of equilibrium, the definitions (\ref{def1}) and (\ref{def2}) have a different meaning. The concept of thermodynamic averaging  is now ill-defined. Instead, the bracket means the need to average over all the available states of the system for the non-equilibrium distributions. Furthermore, the arguments of the Green functions $(x,y)$ are {\it not} usually given as the difference $(x-y)$. For example, non-equilibrium could be caused by transients which make the Green functions
depend upon $(t_x,t_y)$ rather than $(t_x-t_y)$. 

For interacting systems whether in equilibrium or not, one must define and calculate self-energy functions. Again, there are six of them: $\Sigma^{t}$, $\Sigma^{\bar{t}}$, $\Sigma^{<}$, $\Sigma^{>}$, 
$\Sigma^r$ and $\Sigma^a$. The same relationships exist among them as for the Green functions in  (\ref{def1}) and (\ref{def2}), such as
\begin{equation}
\Sigma^r=\Sigma^{t}-\Sigma^{<}=\Sigma^{>}-\Sigma^{\bar{t}}, \:\:\:\:\Sigma^a=\Sigma^{t}-\Sigma^{>}=\Sigma^{<}-\Sigma^{\bar{t}}. 
\end{equation}
The self-energies are incorporated into the Green functions through the use of  Dyson's equations. A useful notation may be introduced which expresses four of the six Green functions as the elements of two-by-two matrices \cite{craig}

\begin{equation}
\widetilde{G}=\left(
\begin{array}{cc}
G^{t} & \pm G^{<}\\
G^{>} & - G^{\bar{t}}
\end{array}\right), \:\:\:\:
\widetilde{\Sigma}=\left(
\begin{array}{cc}
\Sigma^{t} & \pm \Sigma^{<}\\
\Sigma^{>} & - \Sigma^{\bar{t}}
\end{array}\right),
\end{equation}
where the upper signs refer to bosonic case and the lower signs to fermionic case. For systems either in equilibrium or non-equilibrium, Dyson's equation is most easily expressed by using the matrix notation
\begin{equation}
\label{d1}
\widetilde{G}(x,y)=\widetilde{G}^0(x,y)+\int\: d^4x_3\:\int d^4x_4\: \widetilde{G}^0(x,x_3)
\widetilde{\Sigma}(x_3,x_4)\widetilde{G}(x_4,y),
\end{equation}
where the superscript ``0'' on the Green functions means to use those for {\it noninteracting} system.   This equation appears quite formidable; however, some simple expressions may  be obtained for the respective Green functions. It is useful to notice that Dyson's equation can be written in an alternate form, instead of  (\ref{d1}), with $\widetilde{G}^0$ on the right in the interaction terms,
\begin{equation}
\label{d2}
\widetilde{G}(x,y)=\widetilde{G}^0(x,y)+\int\: d^4x_3\:\int d^4x_4\: \widetilde{G}(x,x_3)
\widetilde{\Sigma}(x_3,x_4)\widetilde{G}^0(x_4,y).
\end{equation}
Equations. (\ref{d1}) and (\ref{d2}) are the starting points to derive the quantum Boltzmann equations
describing the temporal evolution of the CP-violating particle density asymmetries.

\begin{flushleft}
{\bf  3. QBE  for bosonic particle density asymmetry}
\end{flushleft}

From now on we will adopt the general method of deriving the QBE's provided by 
Kadanoff and Baym \cite{kb}. In this section our goal is  to find the QBE   for the  generic bosonic  CP-violating current
\begin{equation}
\langle J_\phi^\mu(x) \rangle \equiv i \langle \phi^{\dagger}(x)\stackrel{\leftrightarrow}{\partial}_x^\mu  \phi(x)\rangle\equiv \left[ n_\phi(x), \vec{J}_\phi(x)\right].
\end{equation}
The zero-component of this current $n_\phi$ represents the number density of particles minus the number density of antiparticles and is therefore the quantity which enters the diffusion equations of supersymmetric electroweak baryogenesis. 

Since  the CP-violating current can be expressed in terms of the Green function
$G^{<}_{\phi}(x,y)$ as
\begin{equation}
\label{c1}
\langle J_\phi^\mu(x) \rangle= - \left. \left(\partial_x^\mu - \partial_y^\mu\right) G^{<}_{\phi}(x,y)\right|_{x=y},
\end{equation}
 the problem is reduced to find the QBE for the interacting  Green function $G^{<}_{\phi}(x,y)$ when the system is not in equilibrium. This equation  can be found from (\ref{d1}) by operating  by
$\left(\stackrel{\rightarrow}{\Box}_x+m^2\right)$ on both sides of the equation. Here $m$ represents the bare  mass term of the field $\phi$. On the right-hand  side, this operator acts only on $\widetilde{G}_\phi^0$
\begin{equation}
\label{f1}
\left(\stackrel{\rightarrow}{\Box}_x+m^2\right)\widetilde{G}_\phi(x,y)=\delta^{(4)}(x,y)\widetilde{I}_4+
\int\:d^4 x_3 \widetilde{\Sigma}_\phi(x,x_3)\widetilde{G}_\phi(x_3,y),
\end{equation}
where $I$ is the identity matrix. It is useful to also have an equation of motion for the other variable $y$. This is obtained from (\ref{d2}) by operating  by
$\left(\stackrel{\leftarrow}{\Box}_y+m^2\right)$ on both sides of the equation. We obtain
\begin{equation}
\label{f2}
\widetilde{G}_\phi(x,y)\left(\stackrel{\leftarrow}{\Box}_y+m^2\right)=\delta^{(4)}(x,y)\widetilde{I}_4+
\int\:d^4 x_3 \widetilde{G}_\phi(x,x_3)\widetilde{\Sigma}_\phi(x_3,y).
\end{equation}
The two equations (\ref{f1}) and (\ref{f2}) are the starting point for the derivation of the QBE for the particle asymmetries. Let us extract from  (\ref{f1}) and (\ref{f2}) the equations of motions for the Green function $G^{<}_{\phi}(x,y)$
\begin{eqnarray}
\left(\stackrel{\rightarrow}{\Box}_x+m^2\right)G^{<}_{\phi}(x,y)&=&
\int\:d^4 x_3\left[ \Sigma^{t}_{\phi}(x,x_3)G^{<}_{\phi}(x_3,y)-\Sigma^{<}_{\phi}(x,x_3)G^{\bar{t}}_{\phi}(x_3,y)\right],\\
G^{<}_{\phi}(x,y)\left(\stackrel{\leftarrow}{\Box}_y+m^2\right)&=&
\int\:d^4 x_3\left[ G^{t}_{\phi}(x,x_3)\Sigma^{<}_{\phi}(x_3,y)-G^{<}_{\phi}(x,x_3)\Sigma^{\bar{t}}_{\phi}(x_3,y)\right].
\end{eqnarray}
If we now substract the  two equations and make the identification  $x=y$, the left-hand side is given by 
\begin{equation}
\left. \partial_\mu^x\left[\left(\partial_x^\mu-\partial_y^\mu\right) G^{<}_{\phi}(x,y)\right]\right|_{x=y}=
-\frac{\partial J_\phi^\mu(X)}{\partial X^\mu}=-\left(\frac{\partial n_\phi}{\partial T}+\stackrel{\rightarrow}{\nabla}
\cdot\vec{j}_\phi\right),
\end{equation}
and the QBE for the particle density asymmetry is therefore obtained to be
\begin{equation}
\label{con}
\frac{\partial n_\phi(X)}{\partial T}+\stackrel{\rightarrow}{\nabla}\cdot 
\vec{j}_\phi(X)=-\left. \int\:d^4 x_3\left[\Sigma^{t}_{\phi} G^{<}_{\phi}-\Sigma^{<}_{\phi} G^{\bar{t}}_{\phi}-G^{t}_{\phi} \Sigma^{<}_{\phi}-G^{<}_{\phi} \Sigma^{\bar{t}}_{\phi}\right]\right|_{x=y},
\end{equation}
where we have defined the  center-of-mass coordinate system
\begin{equation}
\label{dd}
X=(T,\vec{X})=\frac{1}{2}(x+y),\:\:\:\: (t,\vec{r})=x-y.
\end{equation}
Notice  that $T$ now means the center-of-mass time and not temperature. The identification $x=y$ in Eq. (\ref{con}) is therefore equivalent to require $t=\vec{r}=0$. 

In order to examine the ``scattering term'' on the right-hand side of Eq. (\ref{con}), the first step is to restore all the variable arguments. Setting  $x=y$ in the original notation of $\Sigma_\phi(x,x_3) G_\phi(x_3,y)$ gives  $(X,x_3)(x_3,X)$ for the pair of arguments
\begin{eqnarray}
\label{s}
\frac{\partial n_\phi(X)}{\partial T}+\stackrel{\rightarrow}{\nabla}\cdot 
\vec{j}_\phi(X)=&-&\int\:d^4 x_3\left[\Sigma^{t}_{\phi}(X,x_3) G^{<}_{\phi}(x_3,X)-\Sigma^{<}_{\phi}(X,x_3) G^{\bar{t}}_{\phi}(x_3,X)\right. \nonumber\\
&+&\left. G^{t}_{\phi} (X,x_3)\Sigma^{<}_{\phi}(x_3,X)-G^{<}_{\phi}(X,x_3) \Sigma^{\bar{t}}_{\phi}(x_3,X)\right].
\end{eqnarray}
The next step is to employ the definitions in (\ref{def1}) to express the time-ordered functions $G^{t}_{\phi}$, $G^{\bar{t}}_{\phi}$, $\Sigma ^t_\phi$, and $\Sigma^{\bar{t}}_{\phi}$ in terms of $G^{<}_{\phi}$, $G^{>}_{\phi}$, 
 $\Sigma^{<}_{\phi}$ and  $G^{>}_{\phi}$.  Then the time integrals are separated into whether
$t_3>T$ or $t_3<T$ and the right-hand side of Eq. (\ref{s}) reads
\begin{eqnarray}
&=&-\int\: d^4 x_3\:\left\{\theta(T-t_3)\left[\Sigma^{>}_{\phi} G^{<}_{\phi}+G^{<}_{\phi}\Sigma^{>}_{\phi}-
\Sigma^{<}_{\phi} G^{>}_{\phi}-G^{>}_{\phi}\Sigma^{<}_{\phi}\right]\right.\nonumber\\
&+&\left. \theta(t_3-T)\left[\Sigma^{<}_{\phi} G^{<}_{\phi}+G^{<}_{\phi}\Sigma^{<}_{\phi}-
\Sigma^{<}_{\phi} G^{<}_{\phi}-G^{<}_{\phi}\Sigma^{<}_{\phi}\right]\right\}.
\end{eqnarray}
The term with $t_3>T$ all cancel, leaving $T>t_3$.  Rearranging these terms gives \cite{riotto97}
\begin{eqnarray}
\label{aaa}
& &\frac{\partial n_\phi(X)}{\partial T}+\stackrel{\rightarrow}{\nabla}\cdot 
\vec{j}_\phi(X)=-\int\: d^3 \vec{x}_3\:\int_{-\infty}^{T}\: dt_3\left[\Sigma^{>}_{\phi}(X,x_3) G^{<}_{\phi}(x_3,X)\right.\nonumber\\
&-&\left. G^{>}_{\phi}(X,x_3) \Sigma^{<}_{\phi}(x_3,X)
+ G^{<}_{\phi}(X,x_3)\Sigma^{>}_{\phi}(x_3,X)-\Sigma^{<}_{\phi}(X,x_3) G^{>}_{\phi}(x_3,X)\right].
\end{eqnarray}
This equation is the QBE for the particle density asymmetry and it can be explicitly checked that, in the particular case in which interactions conserve the number of particles and the latter are neither created nor destroyed, the number asymmetry $n_\phi$ is conserved and  obeys the equation of continuity $\partial n_\phi/\partial T+\stackrel{\rightarrow}{\nabla}\cdot 
\vec{j}_\phi=0$. 
During the production of the baryon asymmetry, however, particle asymmetries are not conserved. This occurs because the 
interactions themselves do not conserve the  particle number asymmetries and  there is some source of CP-violation in the system.  
The right-hand side of Eq. (\ref{aaa}), through the general form of the self-energy $\Sigma_\phi$,  contains all the information necessary to describe the temporal evolution of the particle density asymmetries:  particle number
changing reactions and CP-violating source terms,  which will  pop out from the corresponding self-energy $\Sigma_{{\rm CP}}$. If the interactions of the system do not violate CP,   there will be no CP-violating sources and the final baryon asymmetry produced during supersymmetric baryogenesis will be vanishing. 

As we shall see, the kinetic  Eq. (\ref{aaa}) has an obvious interpretation in terms of gain and loss processes.   
What is unusual, however,   is the presence of the integral over the time: the equation is manifestly non-Markovian. Only  the assumption that the relaxation time scale of the particle asymmetry is much longer than the time scale of the non-local kernels leads to a Markovian description. A further approximation, {\it i.e.} taking the upper limit of the 
time integral to $T\rightarrow \infty$,  leads to the familiar Boltzmann  equation. 
The physical interpretation of the integral over the past history of the system is straightforward: it leads to the typical ``memory'' effects which are observed in quantum transport theory \cite{dan,henning}. In  the classical kinetic theory the ``scattering term'' does not include any integral over the past history of the system which is equivalent to assume that any collision in the plasma  does not depend upon the previous ones. On the contrary,   
quantum distributions posses strong memory effects and the thermalization rate obtained from the quantum transport theory may be substantially longer than the one obtained from the classical kinetic theory. As shown in \cite{riotto97}, 
memory effects play a fundamental role in the determination of the CP-violating sources which fuel baryogenesis when transport properties allow the CP-violating charges to diffuse in front of the bubble wall separating the broken from the unbroken phase at the electroweak phase transition. 

Notice that so far we have not made any approximation and the computation is therefore  valid for all
shapes and sizes of the bubble wall expanding in the thermal bath during a
 first-order electroweak phase transition.
\newpage

\begin{flushleft}
{\bf  4. QBE for fermionic particle density asymmetry}
\end{flushleft}

The generic fermionic CP-violating current reads
\begin{equation}
\langle J_\psi^\mu(x) \rangle \equiv  \langle \bar{\psi}(x)\gamma^\mu \psi(x)\rangle\equiv \left[ n_\psi(x), \vec{J}_\psi(x)\right],
\end{equation}
where $\psi$ indicates  a Dirac fermion  and $\gamma^\mu$ represent the usual Dirac matrices. Again, the zero-component of this current $n_\psi$ represents the number density of particles minus the number density of antiparticles and is therefore the relevant quantity  for the diffusion equations of supersymmetric electroweak baryogenesis.

We want to find a couple of  equations of motion for the interacting fermionic Green function $\widetilde{G}_\psi(x,y)$ when the system is not in equilibrium. Such  equations  may be found  by applying  the operators $\left(i\stackrel{\rightarrow}{\not  \partial}_x -M\right)$
and $\left(i\stackrel{\leftarrow}{\not  \partial}_y +M\right)$  on both sides of  Eqs. (\ref{d1}) and (\ref{d2}), respectively. Here $M$ represents the bare mass term of the fermion $\psi$. We find
\begin{eqnarray}
\label{c}
\left(i\stackrel{\rightarrow}{\not  \partial}_x -M\right)\widetilde{G}_\psi(x,y)&=&\delta^{(4)}(x,y)\widetilde{I}_4+
\int\:d^4 x_3 \widetilde{\Sigma}_\psi(x,x_3)\widetilde{G}_\psi(x_3,y),\\
\widetilde{G}_\psi(x,y)\left(i\stackrel{\leftarrow}{\not  \partial}_y +M\right)&=& -\delta^{(4)}(x,y)\widetilde{I}_4
-\int\:d^4 x_3 \widetilde{G}_\psi(x,x_3)\widetilde{\Sigma}_\psi(x_3,y).
\end{eqnarray}
We can  now  take the trace over the spinorial indeces of  both sides of the equations, sum up  the two equations above  and finally extract the equation of motion for the Green function $G^{>}_{\psi}$
\begin{eqnarray}
\label{v}
{\rm Tr} \left\{\left[i\stackrel{\rightarrow}{\not  \partial}_x + i\stackrel{\leftarrow}{\not  \partial}_y\right]
G^{>}_{\psi}(x,y)\right\}&=& \int\:d^4 x_3\:{\rm Tr}\left[\Sigma^{>}_\psi(x,x_3)G^t_\psi(x_3,y)-\Sigma^{\bar{t}}_\psi(x,x_3)G^{>}_\psi(x_3,y)\right.\nonumber\\
&-&
\left. G^{>}_\psi(x,x_3)\Sigma^t_\psi(x_3,y)+G^{\bar{t}}_\psi(x,x_3)\Sigma^{>}_\psi(x_3,y)\right].
\end{eqnarray}
Making use of the center-of-mass coordinate system,  we can work out the left-hand side of Eq.  (\ref{v})
\begin{eqnarray}
& &\left.{\rm Tr}\left[i\stackrel{\rightarrow}{\not  \partial}_x G^{>}_\psi(T,\vec{X},t,\vec{r})+
G^{>}_\psi(T,\vec{X},t,\vec{r})i\stackrel{\leftarrow}{\not  \partial}_y\right]\right|_{t=\vec{r}=0}\nonumber\\
&=&\left. i\left(  \partial^x_\mu + \partial^y_\mu\right) i \langle
\bar{\psi}\gamma^\mu \psi\rangle\right|_{t=\vec{r}=0}\nonumber\\
&=&-\frac{\partial}{\partial X^\mu} \langle
\bar{\psi}(X)\gamma^\mu \psi(X)\rangle\nonumber\\
&=&-\frac{\partial}{\partial X^\mu} J^\mu_\psi.
\end{eqnarray}
The next step is to employ the definitions in (\ref{def2}) to express the time-ordered functions $G^{t}_{\psi}$, $G^{\bar{t}}_{\psi}$, $\Sigma ^t_\psi$, and $\Sigma^{\bar{t}}_{\psi}$ in terms of $G^{<}_{\psi}$, $G^{>}_{\psi}$, 
 $\Sigma^{<}_{\psi}$ and  $G^{>}_{\psi}$. The computation goes along the same lines as the analysis made in the previous section and we get \cite{riotto97}
\begin{eqnarray}
\label{b}
& & \frac{\partial n_\psi(X)}{\partial T}+\stackrel{\rightarrow}{\nabla}\cdot 
\vec{j}_\psi(X)=\int\: d^3 \vec{x}_3\:\int_{-\infty}^{T}\: dt_3\:{\rm Tr}\left[\Sigma^{>}_{\psi}(X,x_3) G^{<}_{\psi}(x_3,X)\right.\nonumber\\  &-& \left. G^{>}_{\psi}(X,x_3) \Sigma^{<}_{\psi}(x_3,X)     
+ G^{<}_{\psi}(X,x_3)\Sigma^{>}_{\psi}(x_3,X)-\Sigma^{<}_{\psi}(X,x_3) G^{>}_{\psi}(x_3,X)\right].
\end{eqnarray}
This is the ``diffusion'' equation describing the temporal evolution of a generic fermionic number asymmetry $n_\psi$. As for the bosonic case, all the information regarding particle number violating interactions and CP-violating sources  are  stored in the self-energy $\Sigma_\psi$.

\begin{flushleft}
{\bf  5. The  QBE for the right-handed stop number}
\end{flushleft}
 
As we mentioned  in the introduction, a strongly first order electroweak
phase transition
can  be achieved in the presence of a top squark
lighter than the top quark~\cite{r1,r2}.
In order to naturally
suppress its contribution to the parameter $\Delta\rho$ and hence
preserve a good agreement with the precision measurements at LEP,
it should be mainly right-handed. This can be achieved if the left-handed stop soft supersymmetry breaking mass $\widetilde{m}_{\widetilde{t}_L}$
is much larger than $M_Z$. Under this assumption, only the right-handed stops contribute to the 
 axial stop charge. 
The right-handed stop current
$J^\mu_{\widetilde{t}_R}$ associated to the right-handed stop $\widetilde{t}_R$
is given by
\begin{equation}
J_{\widetilde{t}_R}^\mu=i\left(\widetilde{t}^*_R\stackrel{\leftrightarrow}
{\partial^\mu}\widetilde{t}_R\right).
\end{equation}

The self-energy of the right-handed stop contains the information about all the  sources which are responsible for changing $n_{\widetilde{t}_R}$ in the plasma: scattering processes involving the top quark Yukawa coupling, axial top number violation processes, strong sphaleron interactions and the CP-violating source induced by the presence of CP-violating phases in the interactions of the right-handed stop with the  Higgs background.

Eq. (\ref{aaa}) is the QBE describing  the right-handed stop number asymmetry. Solving this equation  represents an Herculean task  since it is     integral  and nonlinear.  This happens because the self-energy functions $\Sigma^{>}$ and $\Sigma^{<}$ are also functions of the full non-equilibrium Green functions
of other degrees of freedom of the system.  We can make some progress, though. 
Since we know that $n_{\widetilde{t}_R}$ identically vanishes if   there is no CP-violating source in the QBE, {\it i.e.}  in absence of the  Higgs configuration describing the bubble wall profile, we first decompose the generic Green function $\widetilde{G}$ and self-energy function $\widetilde{\Sigma}$ as
\begin{equation}
\widetilde{G}=\widetilde{G}^0+\delta \widetilde{G},\:\:\:
\widetilde{\Sigma}=\widetilde{\Sigma}^0+\delta \widetilde{\Sigma},
\end{equation}
where $\widetilde{G}^0$ and $\widetilde{\Sigma}^0$ represent 
the {\it fully interacting} equilibrium Green functions  and self-energy functions 
in the {\it  unbroken} phase (that is  in the absence of the Higgs profile). This approximation amounts to retaining only the first-order linear  response
in the ``Higgs insertion expansion'' around the symmetric phase $\langle H_i^0(x)\rangle=v_i(x)=0$ $(i=1,2)$. This is certainly a good approximation for the case under investigation. We will return to this point later.

If we now linearize Eqs. (\ref{d1}) and (\ref{d2})  and repeat the procedure described in section 3, we obtain 
\begin{eqnarray}
\label{aaaa}
\frac{\partial n_{\widetilde{t}_R}}{\partial T}&+&\stackrel{\rightarrow}{\nabla}\cdot 
\vec{j}_{\widetilde{t}_R}=-\int d^3 \vec{x}_3\int_{-\infty}^{T} dt_3\left[\delta\Sigma^{>}_{\widetilde{t}_R} G^{0,<}_{\widetilde{t}_R}- G^{0,>}_{\widetilde{t}_R} \delta\Sigma^{<}_{\widetilde{t}_R}\right.\nonumber\\
&+& \left. G^{0,<}_{\widetilde{t}_R}\delta\Sigma^{>}_{\widetilde{t}_R}-\delta
\Sigma^{<}_{\widetilde{t}_R} G^{0,>}_{\widetilde{t}_R}+
\Sigma^{0,>}_{\widetilde{t}_R} \delta G^{<}_{\widetilde{t}_R}
\right.\nonumber\\
&-&\left. \delta G^{>}_{\widetilde{t}_R} \Sigma^{0,<}_{\widetilde{t}_R}
+ \delta G^{<}_{\widetilde{t}_R}\Sigma^{0,>}_{\widetilde{t}_R}-
\Sigma^{0,<}_{\widetilde{t}_R} \delta G^{>}_{\widetilde{t}_R}\right],
\end{eqnarray}
where we have used the fact that $n^0_{\widetilde{t}_R}=\vec{j}^0_{\widetilde{t}_R}=0$ and therefore the right-handed stop number asymmetry and current are associated to the Green function $\delta G^{<}_{\widetilde{t}_R}$. 

In the spirit of the diagrammatic approach in the Higgs insertion expansion, we can now  write the shift in the self-energy as
\begin{equation}
\delta\Sigma_{\widetilde{t}_R}=\delta\Sigma^{{\rm CP}}_{\widetilde{t}_R}+
\delta\Sigma^{{\rm int}}_{\widetilde{t}_R}+\cdots,
\end{equation}
where $\delta\Sigma^{{\rm CP}}_{\widetilde{t}_R}$ is the part of the self-energy responsible for the appearance of the CP-violating source and
$\delta\Sigma^{{\rm int}}_{\widetilde{t}_R}$ accounts the interactions   which change the small particle number asymmetry of the right-handed stop originated at the passage of the Higgs profile through a given region of space. 
Here the dots represent other terms in the self-energy describing the interactions which do not change the small particle number asymmetry of the right-handed stop (elastic scatterings) -- they will not give any contribution to the right-hand side of the QBE. 

Eq. (\ref{aaaa}) contains in the right-hand side  generic expressions like 
\begin{equation}
\delta\Sigma^{{\rm CP}}_{\widetilde{t}_R}G^0_{\widetilde{t}_R}+
\delta\Sigma^{{\rm int}}_{\widetilde{t}_R}G^0_{\widetilde{t}_R}+
\Sigma^0_{\widetilde{t}_R}\delta G_{\widetilde{t}_R}
\end{equation}
and
\begin{equation}
G^0_{\widetilde{t}_R}\delta\Sigma^{{\rm CP}}_{\widetilde{t}_R}+
G^0_{\widetilde{t}_R}\delta\Sigma^{{\rm int}}_{\widetilde{t}_R}+
\delta G_{\widetilde{t}_R} \Sigma^0_{\widetilde{t}_R}.
\end{equation}
Let us first analyze the CP-violating source for the right-handed stop number induced by a novanishing $\delta\Sigma^{{\rm CP}}_{\widetilde{t}_R}$. 

\begin{flushleft}
{\it --The  CP-violating source  for the right-handed stop number--}
\end{flushleft}

The interaction which is responsible for the CP-violating source right-handed stop
$\widetilde{t}_R$, the left-handed stop $\widetilde{t}_L$ and the two neutral Higgses 
$H_{1,2}^0$ is given by 
\begin{equation}
\label{interaction}
{\cal L}=h_t \widetilde{t}_L\left(A_t H_2^0-\mu^* H_1^0\right)\widetilde{t}_R^*+ {\rm h.c.}
\end{equation}
Here the soft trilinear term $A_t$ and the supersymmeric mass term $\mu$ are meant to be complex parameters so that ${\rm Im}(A_t\mu)$ is nonvanishing.

At the lowest level of perturbation, the interactions (\ref{interaction}) induce a contribution to 
the  self-energy of the form \cite{riotto97}
\begin{equation}
\label{q}
\delta \Sigma^{{\rm CP},>}_{\widetilde{t}_R}(x,y)=g^{{\rm CP}}(x,y)G^{>}_{\widetilde{t}_L}(x,y), \:\:\:\:
\delta\Sigma^{{\rm CP},<}(x,y)=g^{{\rm CP}}(x,y)G^{<}_{\widetilde{t}_L}(x,y).
\end{equation}
Here 
\begin{equation}
\label{u}
g_{{\rm CP}}(x,y)=h_t^2 \left[A_t^* v_2(x)-\mu v_1(x)\right] \left[A_t v_2(y)-\mu^* v_1(y)\right]
\end{equation}
and we can safely approximate the exact left-handed stop Green functions
$G_{\widetilde{t}_L}$ with the corresponding fully interacting equilibrium Green functions in the unbroken phase $G^{0}_{\widetilde{t}_L}$. 
This is because any departure from thermal equilibrium distribution
functions
is caused at a given point by the passage of the wall and, therefore,
is  ${\cal O}(v_{\omega})$.  Since we will show that the source  is
already linear in $v_{\omega}$,
working with thermal {\it equilibrium} Green 
functions in the unbroken phase
amounts to ignoring terms of higher order in
$v_{\omega}$. This is 
 as accurate as the bubble wall is moving slowly in
the plasma. 

If we now insert the expressions (\ref{q}) and  (\ref{u}) into the QBE 
(\ref{aaaa}), we get the right-handed stop CP-violating source \cite{riotto97}
\begin{eqnarray}
{\cal S}_{\widetilde{t}_R}&=&- 2 i \int\: d^3 \vec{x}_3\:\int_{-\infty}^{T}\: dt_3\left[g_{{\rm CP}}(X,x_3)-g_{{\rm CP}}(x_3,X)\right]\nonumber\\
&\times& {\rm Im}\left[G^{0,>}_{\widetilde{t}_L}(X,x_3) G^{0,<}_{\widetilde{t}_R}(x_3,X)\right]+\cdots\nonumber\\
&=& 4\:h_t^2 \int\: d^3 \vec{x}_3\:\int_{-\infty}^{T}\: dt_3 \:{\rm Im}\left( A_t\mu\right)\left[v_2(X)v_1(x_3)-
v_2(x_3) v_1(X)\right]\nonumber\\
&\times& {\rm Im}\left[G^{0,>}_{\widetilde{t}_L}(X,x_3) G^{0,<}_{\widetilde{t}_R}(x_3,X)\right].
\end{eqnarray}
${\cal S}_{\widetilde{t}_R}$  vanishes if the relative phase of $A_t\mu$ is zero and if the ratio ${\rm tan}\beta(x)\equiv v_2(x)/v_1(x)$ is a constant in the interior of the bubble wall.    

Notice that the source is built up integrating over all the history of the system. It is exactly this  ``memory effect'' that is responsible for the enhancement of the final baryon asymmetry. 
Furthermore,  the source is constructed  when the  right-handed stops pass across the wall, they first scatter off the wall and are transformed into left-handed stops; the latter subsequently suffer another scattering off the wall and are converted again into right-handed stops. If  CP-violation is taking place in both interactions, a nonvanishing 
CP-violating source ${\cal S}_{\widetilde{t}_R}$ pops out from thermal bath. 

In order to deal with analytic expressions, we can work out
the thick wall limit and simplify the expressions obtained above
by performing a derivative expansion
\begin{equation}
\label{expansion}
v_i(x_3)= \sum_{n=0}^{\infty}\frac{1}{n!}\; \frac{\partial^n}
{\partial (X^\mu)^n} v_i(X)\left(x^\mu_3-X^\mu\right)^n .
\end{equation}
The term
with no
derivatives vanishes in the expansion (\ref{expansion}),
$v_2(X)v_1(X)-v_1(X)v_2(X)= 0$, which means that the static
term in the derivative expansion  does not contribute
to the source  ${\cal S}_{\widetilde{t}_R}$.
For a smooth Higgs profile, the derivatives with
respect to the time coordinate and $n>1$ are associated with higher
powers of $v_{\omega}/L_{\omega}$,  where $v_{\omega}$ and $L_{\omega}$ are the velocity and the width of the bubble wall, respectively.  Since the typical time scale of the processes giving rise to the source is given by the thermalization time of the stops $1/\Gamma_{\widetilde{t}}$, 
the approximation is good for values of
$L_{\omega}\Gamma_{\stop}/v_{\omega} \gg 1$.
In other words, this expansion is valid only when the mean free path of the stops in the plasma 
 is smaller than the scale of variation of the Higgs
background determined by the wall thickness, $L_{\omega}$,
and the wall velocity $v_{\omega}$. A detailed 
computation of the   thermalization rate of the  right-handed stop from the imaginary
 part of the two-point Green function has been recently performed in \cite{therm} by making use of improved propagators and including
 resummation of hard thermal loops\footnote{The left-handed stop is usually considered to be much heavier than $T$ and its decay width corresponds to the one in the present vacuum.}. The thermalization rate has been  computed exactly at the
 one-loop level in the high temperature approximation as a function of the plasma right-handed stop mass 
 $m_{\widetilde{t}_R}(T)$ and an  estimate for the magnitude of the
 two-loop contributions which dominate the rate for small $m_{\widetilde{t}_R}(T)$ was also given. 
If  $m_{\widetilde{t}_R}(T)\gsim  T$, the thermalization  is dictated by the one-loop thermal decay rate which can be larger than $T$ \cite{therm}\footnote{For smaller values of   $m_{\widetilde{t}_R}(T)$, when the 
 thermalization is dominated by two-loop effects ({\it i.e.} scattering),   $\Gamma_{\widetilde{t}_R}$ may be as large as $10^{-3} T$ \cite{therm}.}.

The  term corresponding to  $n=1$  in the expansion (\ref{expansion})  gives a contribution to the source proportional to the function 
\begin{equation}
v_1(X)\partial_X^\mu v_2(X)- v_2(X)\partial_X^\mu v_1(X)
\equiv v^2(X) \partial_X^\mu\beta(X),
\end{equation}
which  should vanish smoothly for values of $X$ outside the
bubble wall. Here  we have denoted $v^2\equiv v_1^2+ v_2^2$. 
Since the variation of the Higgs fields is due to  the
expansion of the bubble wall through the thermal bath,
the source ${\cal S}_{\widetilde{t}_R}$
will be linear in $v_{\omega}$. The corresponding contribution tot he CP-violating source reads \cite{riotto97}
\begin{equation}
\label{source1}
{\cal S}_{\widetilde{t}_R}(X)=h_t^2\:{\rm Im}\left( A_t\mu\right) v^2(X)\dot{\beta}(X)\:{\cal I}_{\widetilde{t}_R},
\end{equation}
where $\dot{\beta}(X)\equiv d\beta(X)/dt_X$, 
\begin{eqnarray}
{\cal I}_{\widetilde{t}_R} &=& \int_0^\infty dk \frac{k^2}{2 \pi^2  \;
\omega_{\widetilde{t}_L} \; \omega_{\widetilde{t}_R}}   \nonumber\\
&\times& \left[ \left(1 + 2 {\rm Re}(f^0_{\widetilde{t}_L}) \right)
I(\omega_{\widetilde{t}_R},\Gamma_{\widetilde{t}_R},\omega_{\widetilde{t}_L},\Gamma_{\widetilde{t}_L})
+
\left(1 + 2 {\rm Re}(f^0_{\widetilde{t}_R}) \right)
I(\omega_{\widetilde{t}_L},\Gamma_{\widetilde{t}_L},\omega_{\widetilde{t}_R},\Gamma_{\widetilde{t}_R})\right.  \nonumber\\
&-&\left. 
2 \left( {\rm Im}(f^0_{\widetilde{t}_R}) +
{\rm Im}(f^0_{\widetilde{t}_L}) \right) G(\omega_{\widetilde{t}_R},\Gamma_{\widetilde{t}_R},
\omega_{\widetilde{t}_L},\Gamma_{\widetilde{t}_L}) \right]
\end{eqnarray}
and 
$f^0_{\widetilde{t}_R(L)} = 1/\left[\exp\left(\omega_{\widetilde{t}_R(L)}/T + i \Gamma_{\widetilde{t}_R(L)}/T
\right)- 1 \right]$ are the equilibrium distribution functions in the unbroken phase\footnote{To account for interactions with the surrounding
plasma in the unbroken phase, particles must be substituted by quasiparticles, dressed propagators are to be adopted. This means that self-energy corrections ate one and twp-loops to the particle propagators modify the dispersion relations, which becomes $\omega^2=\vec{k}^2+m^2(T)$ with $m(T)$ the plasma mass, and introduce a finite width $\Gamma$.}.

The functions $I$ and $G$ are given by 
\begin{eqnarray}
\label{v1}
I(a,b,c,d) &=& \frac{1}{2}\frac{1}{\left[(a+c)^2 + (b+d)^2 \right]}
\sin\left[ 2{\rm arctan}\frac{a+c}{b+d}\right]\nonumber\\
&+&\frac{1}{2}\frac{1}{\left[(a-c)^2 + (b+d)^2 \right]}
\sin\left[ 2{\rm arctan}\frac{a-c}{b+d}\right],\nonumber\\
G(a,b,c,d)=&-&\frac{1}{2}\frac{1}{\left[(a+c)^2 + (b+d)^2 \right]}
\cos\left[ 2{\rm arctan}\frac{a+c}{b+d}\right]\nonumber\\
&-&\frac{1}{2}\frac{1}{\left[(a-c)^2 + (b+d)^2 \right]}
\cos\left[ 2{\rm arctan}\frac{a-c}{b+d}\right].
\end{eqnarray}
Notice that the function $G(\omega_{\widetilde{t}_R},\Gamma_{\widetilde{t}_R},
\omega_{\widetilde{t}_L},\Gamma_{\widetilde{t}_L})$ has a peak for $\omega_{\widetilde{t}_R}\sim\omega_{\widetilde{t}_L}$.  
This resonant behaviour \cite{noi,riotto97}  is associated to the fact that 
the Higgs background  is
carrying a very low momentum (of order of the inverse of the bubble wall
width $L_\omega$) and to the 
possibility of absorption or emission of Higgs quanta by the
propagating supersymmetric particles. The resonance  can only take place when  the left-handed stop and the right-handed stop  do not differ too much in mass.
By using the Uncertainty Principle, it is easy to understand that the
width of this resonance is expected
to be proportional to the thermalization rate  of the particles giving rise to
the baryon asymmetry.  
Within the MSSM, however, it is assumed that the left-handed stop mass  $m_ {\widetilde{t}_L}$ is much larger than the temperature $T$ and the resonance  can only happen for
momenta larger than  $m_ {\widetilde{t}_L}$. Such configurations are
exponentially suppressed and do not give any relevant
contribution to the CP-violating source. Nonertheless, if the  electroweak phase transition is enhanced by the presence of some
new degrees of freedom beyond the ones contained in the MSSM, {\it
e.g.} some extra standard model gauge singlets, the resonance effects in the stop sector might be relevant.  

\begin{flushleft}
{\it --The  inelastic interactions of the right-handed stop number--}
\end{flushleft}

Let us now investigate the  other terms in Eqs. (30) and (31) leading to particle number changing interactions.  The inelastic scattering processes which change the right-handed stop number are the ones induced by the top Yukawa coupling and by the axial top number violation inside the Higgs bubble wall (besides the the ones arising from strong sphaleron transitions). For sake of simplicity, let us focus on the   top Yukawa interactions --other interactions may be treated similarly. If we
 assume that the left-handed stop is heavier than the temperature of the plasma and is therefore decoupled from the thermal bath, these interactions read 
\begin{equation}
{\cal L}=h_t \overline{\widetilde{H}_2^0} \:P_L t \:\widetilde{t}_R^* +{\rm h.c.},
\end{equation}
where $P_L$ denotes the left-handed chirality projector operator and  $\widetilde{H}^{0}_2$ and $t$ are the Higgsino and top quark fields, respectively. 

The right-hand side of Eq. (\ref{aaaa}) contains now generic expressions like
\begin{equation}
\label{p}
\delta\Sigma^{{\rm int}}_{\widetilde{t}_R}G^0_{\widetilde{t}_R}+
\Sigma^0_{\widetilde{t}_R}\delta G_{\widetilde{t}_R}-
G^0_{\widetilde{t}_R} \delta\Sigma^{{\rm int}}_{\widetilde{t}_R}-\delta G_{\widetilde{t}_R}\Sigma^0_{\widetilde{t}_R}.
\end{equation}
Since the contribution to $\Sigma^{{\rm int}}_{\widetilde{t}_R}$ induced by   the top Yukawa  quark coupling is symbolically  given by $h_t^2 G_{\widetilde{H}^{0}_2}
G_{t_L}$, the expression (\ref{p}) may be written as
\begin{eqnarray}
\label{l}
&&\delta G_{\widetilde{H}^{0}_2}
G^0_{t_L}G^0_{\widetilde{t}_R}+G^0_{\widetilde{H}^{0}_2}
\delta G_{t_L}G^0_{\widetilde{t}_R}+ G^0_{\widetilde{H}^{0}_2}
G^0_{t_L}\delta G_{\widetilde{t}_R}\nonumber\\
&-& G^0_{\widetilde{t}_R}\delta G_{\widetilde{H}^{0}_2}
G^0_{t_L}- G^0_{\widetilde{t}_R} G^0_{\widetilde{H}^{0}_2}
\delta G_{t_L}-  \delta G_{\widetilde{t}_R} G^0_{\widetilde{H}^{0}_2}
G^0_{t_L}.
\end{eqnarray}
We now show that --under a number of reasonable approximations-- these terms all together lead to the  familiar ``scattering term'' proportional to $\left(n_{\widetilde{t}_R}-n_{t_L}+n_{\widetilde{H}_2^0}\right)$ in the right-hand side of the diffusion equation for the right-handed stop number.

\begin{flushleft}
{\it --Getting the diffusion equation--}
\end{flushleft}

To progress further we need to introduce reasonable approximations based on physical considerations. The main phenomena occuring in the plasma when the Higgs bubble wall is passing through a certain region are characterized by 
three different time (or length) scales: the quantum (microscopic) time $\tau_Q$, the statistical (macroscopic) time $\tau_S$ and the time set by the bubble wall thickness and velocity
$\tau_\omega\sim L_\omega/v_\omega$. The first measures the range of radiative corrections to the Compton wavelength of particles and the second measures the range of interactions among particles. In the more familiar classic kinetic interaction theory, a similar distinction is made between the scattering length and the mean free path. It is a reasonable assumption that the statistical
scale $\tau_S$ is much larger than the quantum one. Furthermore, in the thick wall limit, the following hierarchy holds: $\tau_\omega\gg \tau_S\gg \tau_Q$.
This observation allows us to recast the quantum field theoretical problem into 
a much simpler kinetic theory and, under some further assumptions, to rederive the classical diffusion equations --using well-known techniques from relativistic many-body theory \cite{dan}. Suppose we separate the space-time ``cells''
whose characteristic length scale is intermediate between the statistical and the quantum scales. As the correlation between cells will be negligible by construction, the only interesting case is when the two arguments of a given propagator or self-energy lie in the same cell. In the interior of a single cell, relaxation phenomena are negligible. More concretely, the propagators may be Fourier transformed over a cell. Relaxation phenomena become apparent as we move from cell to cell. This picture breaks down if, for instance, the major contribution to CP-violating sources come from long wavelength particles for which the mean free path becomes comparable to the Compton wavelength. We will return to this point later.

In mathematical language, if $x$ and $y$ are the arguments of a propagator or a self-energy, we Fourier transform with respect to the $r=x-y$, while the absolute coordinate $X=\frac{1}{2}(x+y)$ serves as a cell label. The Fourier
transform reads
\begin{equation}
G(x,y)=\int\:\frac{d^4 p}{(2\pi)^4}\:{\rm e}^{-ip\cdot(x-y)}G\left(
\frac{x+y}{2},p\right)=\int\:\frac{d^4 p}{(2\pi)^4}\:{\rm e}^{-ip\cdot r}G\left(X,p\right),
\end{equation}
where $G(X,p)$ is called the Wigner transform of $G(x,y)$. Similar formulae hold for the self-energies. 

Let us now have a closest look at the third term in Eq. (\ref{l}).
We can rewrite it as
\begin{eqnarray}
\Sigma^0_{\widetilde{t}_R}(X,x_3)\delta G_{\widetilde{t}_R}(x_3,X)&=&
\int\:\frac{d^3 \vec{q}}{(2\pi)^3}\:{\rm e}^{i \vec{q}\cdot(\vec{X}-\vec{x}_3)}
\int\:\frac{d^3 \vec{k}}{(2\pi)^3}\:{\rm e}^{i \vec{k}\cdot(\vec{x}_3-\vec{X})}
\Sigma^0_{\widetilde{t}_R}\left(\frac{\vec{X}+\vec{x}_3}
{2},T,t_3,\vec{q}\right)\nonumber\\
& \times & \delta G_{\widetilde{t}_R}\left(\frac{\vec{X}+\vec{x}_3}{2},T,t_3,\vec{k}\right).
\end{eqnarray}
The integrand will be appreciably different from zero only when $X$ belongs to the same cell as $x_3$, that is $(X+x_3)/2\sim 
X$.  Inserting the above expression into Eq. (\ref{aaaa}), we get
\begin{equation}
-\int_{-\infty}^T\:d t_3\: \int\:\frac{d^3 \vec{k}}{(2\pi)^3}\:
\Sigma^0_{\widetilde{t}_R}\left(X,\vec{q},T-t_3\right)
\delta G_{\widetilde{t}_R}\left(X,\vec{k},t_3-T\right).
\end{equation}
A similar expression is obtained for the last term in (\ref{l})
$\delta G_{\widetilde{t}_R}\Sigma^0_{\widetilde{t}_R}$.

Let us now adopt the quasiparticle Ansatz for the shift in the right-handed stop  Green function
\begin{eqnarray}
\label{ansatz1}
\delta G_{\widetilde{t}_R}^{>}(X,k)&=&
\frac{i\pi}{\omega_{\widetilde{t}_R}}
\left[\left(1+\delta f_{\widetilde{t}_R}(X,\vec{k})\right)
\delta(k^0-\omega_{\widetilde{t}_R})\right.\nonumber\\
&+&\left. \delta \bar{f}_{\widetilde{t}_R}(X,-\vec{k})\delta(k^0+
\omega_{\widetilde{t}_R})\right],\nonumber\\
\delta G_{\widetilde{t}_R}^{<}(X,k)&=&\frac{i\pi}{\omega_{\widetilde{t}_R}}
\left[\delta f_{\widetilde{t}_R}(X,\vec{k})
\delta(k^0-\omega_{\widetilde{t}_R})\right.\nonumber\\
&+&\left. \left(1+\delta \bar{f}_{\widetilde{t}_R}(X,-\vec{k})\right)\delta(k^0+
\omega_{\widetilde{t}_R})\right],
\end{eqnarray}
where $\omega^2_{\widetilde{t}_R}=\vec{k}^2+m^2_{\widetilde{t}_R}(X,T)$ and 
the plasma mass squared $m^2_{\widetilde{t}_R}(X,T)$ is a function of the spacetime coordinate $X$ as well as of the temperature $T$. This Ansatz
 is the logical generalization of the Green function for free fields and incorporate the renormalization effects of the non-trivial spectral density in the plasma. In this sense the Wigner functions $\delta f_{\widetilde{t}_R}$ and $\delta\bar{f}_{\widetilde{t}_R}$ are the quantum kinetic extension of the classical particls phase-space distributions for the right-handed stop and its antistate. Notice,though,  that this Ansatz does not solve the QBE unless one neglects off-shell terms \cite{sp}. This procedure is however justified if $\tau_\omega\gg \tau_S\gg \tau_Q$. 

From this Ansatz we can construct the equation for the right-handed stop number asymmetry $n_{\widetilde{t}_R}$ by inserting (\ref{ansatz1}) into the right-hand side of Eq. (\ref{aaaa}). After some algebra and exploiting the 
symmetry properties of the Green functions, we find that the third and the last term of Eq. (\ref{l}) lead to
\begin{eqnarray}
\label{g}
\frac{\partial n_{\widetilde{t}_R}}{\partial T}+\stackrel{\rightarrow}{\nabla}\cdot 
\vec{j}_{\widetilde{t}_R}&\propto &
-\int_{-\infty}^{T} d t_3\int d\omega \int\frac{d^3 \vec{k}}{(2\pi)^3}
\Gamma_{\widetilde{t}_R}\left(\omega,\vec{k}\right)\nonumber\\
&\times&\cos\left[\left(\omega-\omega_{\widetilde{t}_R}\right)\left(
T-t_3\right)\right]\left(
\delta f_{\widetilde{t}_R}-\delta \bar{f}_{\widetilde{t}_R}\right)(\vec{k},t_3),
\end{eqnarray}
where we have made use of the fact that the equilibrium self-energy
$\Sigma^0$ does not depend on the ``cell'' label $X$ and
\begin{equation}
\label{h}
\Gamma_{\widetilde{t}_R}\left(\omega,\vec{k}\right)=\frac{
\Sigma^{0,>}_{\widetilde{t}_R}\left(\omega,\vec{k}\right)-
\Sigma^{0,<}_{\widetilde{t}_R}\left(\omega,\vec{k}\right)}{i
 \omega_{\widetilde{t}_R}}.
\end{equation}
The kinetic Eq. (\ref{g}) has an obvious interpretation in terms of gain and loss processes which change the number density asymmetry of the right-handed stop. This is because the rate of change of the particle number asymmetry in the plasma is related  to the imaginary part of the retarded self-energy function, Eq. (\ref{h}). 
However the kinetic equation is non-Markovian and  the retarded time integral and the cosine function replace the more familiar energy conserving function present in the classical kinetic equation. Under which conditions can we obtain the more familiar classical Boltzmann equation? Let us suppose first that the relaxation time scale for $n_{\widetilde{t}_R}$ is much longer than the time scale of the non-local kernel. Under this assumption, $\delta f_{\widetilde{t}_R}(t_3)-\delta \bar{f}_{\widetilde{t}_R}(t_3)$ can be replaced by $\delta f_{\widetilde{t}_R}(T)-\delta \bar{f}_{\widetilde{t}_R}(T)$ and taken outside the time integral. This leads to a Markovian description. A further approximation is obtained taking the upper limit of the time integral to $T\rightarrow \infty$. The cosine function becomes an energy conserving delta function and we obtain
\begin{equation}
\frac{\partial n_{\widetilde{t}_R}}{\partial T}+\stackrel{\rightarrow}{\nabla}\cdot 
\vec{j}_{\widetilde{t}_R}\propto -\int\:\frac{d^3 \vec{k}}{(2\pi)^3}
\:\Gamma^{\infty}_{\widetilde{t}_R}\left(\omega_{\widetilde{t}_R},\vec{k}
\right)
\left(\delta f_{\widetilde{t}_R}-\delta \bar{f}_{\widetilde{t}_R}\right)(\vec{k},T).
\end{equation}
This result is the familiar relationship between the relexation time of the 
particle distribution $\Gamma^{\infty}_{\widetilde{t}_R}$ and the damping rate,
which is determined by the contribution to imaginary part of the self energy on shell from the top Yukawa coupling,
$\Gamma_{\widetilde{t}_R}=-i\left(\Sigma^{0,>}_{\widetilde{t}_R}-\Sigma^{0,<}_{\widetilde{t}_R}\right)/\omega_{\widetilde{t}_R}$. 

We can repeat the same procedure for all the other terms present in the expression (\ref{l}). 
Making the further approximation that the relaxation rate induced by 
the top Yukawa coupling is the {\it same}  for the left-handed top, the right-handed stop and the higgsino\footnote{This approximation may turn out to be  very rough for some choice of the parameters because the imaginary part of the two-point self-energy of different particles depend sensitively on their dispersion relations
in the plasma \cite{therm,inprep} and, especially for Majorana fermions, the latter are highly nontrivial \cite{majorana}.}
-- $\Gamma_{{\rm top}}$--  and that,   in the high temperature limit, it 
does  depend on the three-momentum only weakly
, we can take it outside of the momentum integral. Moreover, if we follow the spirit of the usual derivation of Fick's law, we end up with the  familiar diffusion equation for the right-handed stop number asymmetry
\begin{equation}
\frac{\partial n_{\widetilde{t}_R}}{\partial T}-D_{\widetilde{t}_R}\nabla^2
n_{\widetilde{t}_R}= -\Gamma_{{\rm top}}\left(n_{\widetilde{t}_R}-
n_{t_L}+n_{\widetilde{H}_2^0}\right)+ {\cal S}_{\widetilde{t}_R}+\cdots,
\end{equation}
where 
\begin{equation}
n_{\widetilde{t}_R}=\int\:\frac{d^3 \vec{k}}{(2\pi)^3}\left(\delta f_{\widetilde{t}_R}-\delta \bar{f}_{\widetilde{t}_R}\right), 
\end{equation}
$D_{\widetilde{t}_R}$ is the right-handed stop diffusion coefficient and the dots mean that one should also include the axial top number violation processes as well as the strong sphaleron interactions.

To assess whether the Markovian and the Boltzmann approximations are reliable in the framework of supersymmetric electroweak baryogenesis one has to understand the different time scales. The time scale of the kernel is
of the order of $\omega_{\widetilde{t}_R}^{-1}$. Since $\Gamma_{\widetilde{t}_R}\ll \omega_{\widetilde{t}_R}$, the relaxation time scale of the right-handed stop population is longer than the range of the 
kernel and the Markovian approximation is warrented. Furthermore, since the time scale set by the bubble wall profile $\sim L_\omega/v_\omega$ and the 
diffusion time scale $\sim D_{\widetilde{t}_R}/v_\omega^2$ are larger than  the relaxation time, the Boltzmann/diffusion equation is also warrented\footnote{On the other hand, for soft scales and for values of the right-handed SUSY breaking mass term such that  the plasma mass $m_{\widetilde{t}_R}$  is small compared to the temperature, the relaxation time scale becomes comparable to the time scale of the kernel and the Markovian approximation breaks down.}. 

\begin{flushleft}
{\bf  6. The QBE for the Higgsino number}
\end{flushleft}
 
The Higgs fermion  current associated with  neutral
and charged Higgsinos can be written
as
\be
\label{corhiggs}
J^{\mu}_{\widetilde{H}}=\overline{\widetilde{H}}\gamma^\mu \widetilde{H}
\ee
where $\widetilde{H}$ is the Dirac spinor
\be
\label{Dirac}
\widetilde{H}=\left(
\begin{array}{c}
\widetilde{H}_2 \\
\overline{\widetilde{H}}_1
\end{array}
\right)
\ee
and $\widetilde{H_2}=\widetilde{H}_2^0$ ($\widetilde{H}_2^+$),
$\widetilde{H_1}=\widetilde{H}_1^0$ ($\widetilde{H}_1^-$) for
neutral (charged) Higgsinos. The processes in the plasma which  change the Higgsino number are the ones induced by the top Yukawa coupling and by  interactions with the Higgs profile. 
The interactions among the charginos and the charged Higgsinos which are responsible for the CP-violating source in the diffusion equation for the Higgs fermion number read
\begin{equation}
{\cal L}=-g_2\left\{\overline{\widetilde{H}}\left[v_1(x)P_L+{\rm e}^{i\theta_\mu} v_2(x) P_R\right]\widetilde{W}\right\}+{\rm h.c.},
\end{equation}
where $\theta_\mu$ is the phase of the $\mu$-parameter. Analogously, the interactions among the Bino, the $\widetilde{W}_3$-ino and the neutral Higgsinos are
\begin{equation}
 {\cal L}=-\frac{1}{2}\left\{\overline{\widetilde{H}^0}\left[v_1(x)P_L+{\rm e}^{i\theta_\mu} v_2(x) P_R\right]\left(g_2\widetilde{W}_3-g_1\widetilde{B}\right)\right\}+{\rm h.c.}
\end{equation}
We can compute the QBE for the Higgsino number repeating step by step the procedure adopted in the previous section. 
To compute the source for the Higgs fermion number ${\cal S}_{\widetilde{H}}$ 
 we again  perform a  ``Higgs insertion expansion'' around the symmetric phase. At the lowest level of perturbation, the interactions of the charged Higgsino induce 
a contribution to the self-energy of the form (and analogously for the other component 
$\delta\Sigma^{{\rm CP},>}_{\widetilde{H}}$)
\begin{equation}
\label{qf}
\delta\Sigma^{{\rm CP},<}_{\widetilde{H}}(x,y)=g^L_{{\rm CP}}(x,y)P_L G^{0,<}_{\widetilde{W}}(x,y) P_L+
g^R_{{\rm CP}}(x,y)P_R G^{0,<}_{\widetilde{W}}(x,y) P_R,
\end{equation}
where 
\begin{eqnarray}
\label{qs}
g^L_{{\rm CP}}(x,y)&=&g_2^2 v_1(x) v_2(y){\rm e}^{-i\theta_\mu},\nonumber\\
g^R_{{\rm CP}}(x,y)&=&g_2^2 v_1(y) v_2(x){\rm e}^{i\theta_\mu}.
\end{eqnarray}
Again, we have approximated the exact Green function of winos $G_{\widetilde{W}}$ by the equilibrium Green function in the unbroken phase $G^{0}_{\widetilde{W}}$.
Similar formulae hold for the neutral Higgsinos. 

Analogously to the case of right-handed stops, the dispersion
relations of charginos and neutralinos are changed by high
temperature
corrections~\cite{weldon}. Even though fermionic dispersion
relations
are highly nontrivial,  especially when dealing with Majorana fermions \cite{majorana}, relatively simple expressions
for the equilibrium fermionic spectral functions may be given in the limit
in which the damping rate is smaller than the typical self-energy
of the fermionic excitation ~\cite{henning}.  The computation of the source  goes along the same lines of the calculation done in the previous section and it is easy to show that 
the CP-violating source \cite{riotto97} 
\begin{eqnarray}
{\cal S}_{\widetilde{H}}&=&- \int\: d^3\vec{x}_3\:\int_{-\infty}^{T}\: dt_3\:{\rm Tr}\left[\delta\Sigma^{{\rm CP},>}_{\widetilde{H}}(X,x_3) G^{0,<}_{\widetilde{H}}(x_3,X)-
G^{0,>}_{\widetilde{H}}(X,x_3) \delta\Sigma^{{\rm CP},<}_{\widetilde{H}}(x_3,X)\right.\nonumber\\
&+&\left. G^{0,<}_{\widetilde{H}}(X,x_3)\delta\Sigma^{{\rm CP},>}_{\widetilde{H}}(x_3,X)-\delta\Sigma^{{\rm CP},<}_{\widetilde{H}}(X,x_3) G^{0,>}_{\widetilde{H}}(x_3,X)\right],
\end{eqnarray}
containes in the integrand the following function
\begin{equation}
g^L_{{\rm CP}}(X,x_3)+g^R_{{\rm CP}}(X,x_3)-
g^L_{{\rm CP}}(x_3,X)-g^R_{{\rm CP}}(x_3,X)=2i\sin\theta_\mu\left[v_2(X)v_1(x_3)-v_1(X)v_2(x_3)\right],
\end{equation}
which vanishes if  ${\rm Im}(\mu)=0$ and if the $\tan\beta(x)$ is a constant  along the Higgs profile. Performing the "Higgs derivative expansion", we finally get \cite{riotto97}
\begin{equation}
\label{source2}
{\cal S}_{\widetilde{H}}(X) =
{\rm Im}(\mu)\: \left[ v^2(X)\dot{\beta}(X) \right]
\left[ 3 M_2 \; g_2^2 \; {\cal I}^{\widetilde{W}}_{\widetilde{H}}
 +       M_1 \; g_1^2 \; {\cal I}^{\widetilde{B}}_{\widetilde{H}}
\right],
\end{equation}
where
\begin{eqnarray}
{\cal I}^{\widetilde{W}}_{\widetilde{H}} & = & \int_0^\infty dk
\frac{k^2}
{2 \pi^2 
\omega_{\widetilde{H}} \omega_{\widetilde{W}}} \nonumber\\
&\left[ \phantom{\frac{1}{2^2}} \right.&
 \left(1 - 2 {\rm Re}(f^0_{\widetilde{W}}) \right)
I(\omega_{\widetilde{H}},\Gamma_{\widetilde{H}},
\omega_{\widetilde{W}},\Gamma_{\widetilde{W}})+
\left(1 - 2 {\rm Re}(f^0_{\widetilde{H}}) \right)
I(\omega_{\widetilde{W}},
\Gamma_{\widetilde{W}},\omega_{\widetilde{H}},
\Gamma_{\widetilde{H}}) \nonumber\\
&+&
2 \left( {\rm Im}(f^0_{\widetilde{H}}) +
{\rm Im}(f^0_{\widetilde{W}}) \right)
G(\omega_{\widetilde{H}},
\Gamma_{\widetilde{H}},
\omega_{\widetilde{W}},\Gamma_{\widetilde{W}})
\left.\phantom{\frac{1}{2^2}} \right]
\nonumber\\
\end{eqnarray}
and $\omega^2_{\widetilde{H}(\widetilde{W})}=k^2+ |\mu|^2
(M_2^2)$ while $f^0_{\widetilde{H}(\widetilde{W})} =
1/\left[\exp\left(\omega_{\widetilde{H}(\widetilde{W})}/T
+ i \Gamma_{\widetilde{H}(\widetilde{W})}/T \right)
+ 1 \right]$.
The exact computation of the  damping rate of charged and neutral
Higgsinos $\Gamma_{\widetilde{H}}$ will be given elsewhere \cite{inprep}, but it is expected to be of the order of $5\times 10^{-2} T$. 
The Bino contribution may be obtained from the above
expressions by replacing $M_2$ by $M_1$. As  for ${\cal S}_{\widetilde{t}_R}$, the CP-violating source for the Higgs fermion number is enhanced  if   $M_{2}, M_{1}\sim \mu$ and  low momentum particles are transmitted over the distance $L_\omega$. 
This means that    the classical approximation is not  entirely adequate to
describe the quantum interference nature of $CP$-violation and only a quantum approach is suitable for the computation of the building up of the CP-violating sources \cite{riotto97}. 

To find the ``collision terms'' which change the higgsino number inside the Higgs bubble wall one has to repeat the analysis performed previously  for the right-handed
stop number. In particular, one has to adopt the quasiparticle Ansatz for the Higgsinos. The one for charged Higgsinos, for instance, reads
\begin{eqnarray}
\delta G_{\widetilde{H}}^{>}(X,k)&=&
-\frac{i\pi}{\omega_{\widetilde{H}}}\left(\not k + \left|\mu\right|\right)
\left[\left(1+\delta f_{\widetilde{H}}(X,\vec{k})\right)
\delta(k^0-\omega_{\widetilde{H}})\right.\nonumber\\
&+&\left. \delta \bar{f}_{\widetilde{H}}(X,-\vec{k})\delta(k^0+
\omega_{\widetilde{H}})\right],\nonumber\\
\delta G_{\widetilde{H}}^{<}(X,k)&=-&\frac{i\pi}{\omega_{\widetilde{H}}}
\left(\not k + \left|\mu\right|\right)\left[\delta f_{\widetilde{H}}(X,\vec{k})
\delta(k^0-\omega_{\widetilde{H}})\right.\nonumber\\
&+&\left. \left(1+\delta \bar{f}_{\widetilde{H}}(X,-\vec{k})\right)\delta(k^0+
\omega_{\widetilde{H}})\right],
\end{eqnarray}
where we have assumed that the mass term is well approximated by the bare mass term $\left|\mu\right|$. Inserting this Ansatz into the QBE for the fermionic degrees of freedom (\ref{b}) one can recover the usual Boltzmann/diffusion equation for the Higgsino number inside the Higgs bubble wall once the same approximations discussed in the subsection 5.2 have been made. 

\newpage
\begin{flushleft}
{\bf  7. The  baryon number asymmetry}
\end{flushleft}
Once we have computed the CP-violating sources for the right-handed stop number and the Higgsino number and we have shown that in the thick wall limit the particle number changing interactions in the QBE's   reduce to the more familiar Boltzmann/diffusion equations, we are ready to estimate the final baryon asymmetry produced during the electroweak phase transition. 

From now on we will closely follow the approach taken in
ref.~\cite{nelson,noi} where the
reader is referred to for more details.
If the system is near thermal equilibrium and particles interact weakly,
the particle number densities $n_i$ may be expressed as
$n_i=k_i\mu_iT^2/6$ where
$\mu_i$ is the local chemical potential, and $k_i$ are statistical
factors of order of 2 (1) for light bosons (fermions) in thermal
equilibrium, and Boltzmann suppressed for particles heavier than $T$.

The particle densities we
need to include are the left-handed top
doublet $q_L\equiv(t_L+b_L)$,
the right-handed top quark $t_R$, the Higgs particle
$H\equiv(H_1^0, H_2^0, H_1^-, H_2^+)$, and the superpartners
$\widetilde{t}_R$ and $\widetilde{H}$ --remember that the left-handed stop and sbottom are supposed to be heavier than the temperature $T$.
As usual, we shall assume that the supergauge interactions are in
equilibrium.
Under these assumptions the system may be described by the
densities
${\cal Q} = q_L$,
${\cal {\cal T}}=t_R+\widetilde{t}_R$ and ${\cal H}=H+\widetilde{H}$.
Ignoring the curvature of the bubble wall, any quantity becomes a
function of the coordinate
${\bf z}=z_3+v_{\omega}z$, the coordinate normal
to the wall surface, where we assume the bubble wall is moving along the
$z_3$-axis.

Assuming that the rates of the interactions induced by the top Yukawa coupling $\Gamma_{{\rm top}}$ and by the strong sphalerons $\Gamma_{{\rm ss}}$  are so fast that ${\cal Q}/k_q-
{\cal H}/k_{\cal H}- {\cal T}/k_{\cal T}={\cal O}(1/\Gamma_{{\rm top}})$ and
$2{\cal Q}/k_q-{\cal T}/k_{\cal T}+
9({\cal Q}+{\cal T})/k_b={\cal O}(1/\Gamma_{{\rm ss}})$,
one can find the equation governing the Higgs density \cite{nelson}
\begin{equation}
\label{equation}
v_{\omega}{\cal H}^\prime-\overline{D}
{\cal H}^{\prime\prime}+\overline{\Gamma}{\cal H}-
\widetilde{\gamma}=0,
\end{equation}
where the derivatives are now with respect to ${\bf z}$,
$\overline{D}$ is the effective diffusion constant,
$\widetilde{\gamma}$ is an effective source term
in the frame of the
bubble wall and $\overline{\Gamma}$ is the effective decay
constant~\cite{nelson}.
An analytical solution to Eq.~(\ref{equation}) satisfying the
boundary conditions ${\cal H}(\pm\infty)=0$ may be found in the
symmetric
phase (defined by ${\bf z}<0$) using a ${\bf z}$-independent
effective diffusion constant and a step function for the effective decay rate
$\overline{\Gamma}= \widetilde{\Gamma} \theta({\bf z})$. A more realistic
form of $\overline{\Gamma}$ would interpolate smoothly between the
symmetric and the broken phase values. One can check, however,
that the result is insensitive to the specific position of the
step function  inside the bubble wall \cite{noi}. 
The values of $\overline{D}$ and $\overline{\Gamma}$ in
(\ref{equation}) of course depend on the particular
values of supersymmetric parameters. For the
considered range we typically find $\overline{D}\sim 0.8\ {\rm GeV}^{-1}$,
$\overline{\Gamma}\sim 1.7$ GeV \cite{noi}.

The tunneling processes from the symmetric phase to the true minimum in the first order phase transition of the Higgs field in the MSSM has been recently analyzed in \cite{mariano} including the leading two-loop effects. It was shown that  the Higgs profile along the bubbles at the time when the latter  are formed have a typical thickness $L_\omega\sim (20-30)/T$ \footnote{In general, however,  the value of $L_\omega$ when the bubbles are moving through the plasma with some velocity $v_\omega$  is different from the value at bubble nucleation. Indeed, the motion of the bubble wall is determined by two main factors, namely the pressure difference between inside and outside the bubble --leading to the expansion-- and the friction force, proportional to $v_\omega$, accounting for the collisions of the plasma particles off the wall. The equilibrium between these two forces imples a steady state with a final velocity $v_\omega$. If bubbles are rather thick, thermodinamical conditions are established inside the wall and for the latter is no longer possible to loose energy by thermal dissipation. Under these conditions the bubble wall is accelerated until slightly out-of-equilibrium conditions and the friction forces are reestablished.}. The total amount of the baryon asymmetry is proportional to $\Delta\beta$ --the change in the ratio of the Higgs vacuum expectation values
$\beta=v_2/v_1$ from ${\bf z}=0$ to inside the bubble wall. 
This quantity tends to zero for large values of $m_A$,
and takes small values, of order $10^{-2}$ for values of the pseudoscalar mass 
$m_A = 150$--200 GeV \cite{mariano}. 

The dependence of the final baryon asymmetry on  $\widetilde{\gamma}$
is much more delicate. The solution of Eq. (\ref{equation})
for ${\bf z} < 0$ is
\begin{equation}
\label{higgs1}
{\cal H}({\bf z})={\cal A}\:{\rm e}^{{\bf z}v_{\omega}/\overline{D}},
\end{equation}
and for ${\bf z} >0$ is 
\begin{eqnarray}
\label{higgs3}
{\cal H}({\bf z}) & = & \left( {\cal B}_{+} -
\frac{1}{\overline{D}(\lambda_+  - \lambda_-)}
\int_0^{{\bf z}} du \widetilde \gamma(u) e^{-\lambda_+ u} \right)
e^{\lambda_{+} {\bf z}}
\nonumber\\
&+& \left( {\cal B}_{-} -
\frac{1}{\overline{D}(\lambda_-  - \lambda_+)}
\int_0^{\bf{z}} du \widetilde \gamma(u) e^{-\lambda_- u} \right)
e^{\lambda_{-} {\bf z}}.
\end{eqnarray}
where
\begin{equation}
\lambda_{\pm} = \frac{ v_{\omega} \pm
\sqrt{v_{\omega}^2 + 4 \widetilde{\Gamma}
\overline{D}}}{2 \overline{D}},
\end{equation}
and $\widetilde \gamma({\bf z})$ is the total CP-violating current resulting
from the sum of the right-handed
stop and Higgsino contributions.
Imposing the continuity of ${\cal{H}}$ and
${\cal{H}}'$ at the boundaries, we find
\begin{equation}
\label{higgs2}
{\cal A}= {\cal B}_{+}\left(1-\frac{\lambda_-}{\lambda_+}\right)=
{\cal B}_{-}\left(\frac{\lambda_+}{\lambda_-}-1\right)=
\frac{1}{\overline{D} \; \lambda_{+}} \int_0^{\infty} du\;
\widetilde \gamma(u)
e^{-\lambda_+ u}.
\end{equation}
From the form of the above equations one can see that CP-violating
densities are non zero for a time $t\sim \overline{D}/ v_{\omega}^2$
and the assumptions leading to the analytical
form of ${\cal H}({\bf z})$ --from Eq. (64)--are valid
provided that the interaction rates $\Gamma_{{\rm top}}$ and 
$\Gamma_{{\rm ss}}$  are larger than $v_{\omega}^2/\overline{D}$ \cite{nelson,noi}.

The equation governing
the baryon asymmetry $n_B$ is given by~\cite{nelson}
\begin{equation}
\label{bau}
D_q n_B^{\prime\prime}-v_{\omega} n_B^\prime-
\theta(-{\bf z})n_f\Gamma_{{\rm ws}}n_L=0,
\end{equation}
where $\Gamma_{{\rm ws}}=6\kappa\alpha_w^4T$ is the
weak sphaleron
rate ($\kappa\simeq 1$)~\cite{SphalRate} (the correct value of $\kappa$ is at present subject of debate),
and $n_L$ is the total number density of
left-handed weak doublet fermions, $n_f=3$ is the number of
families and
we have assumed that the baryon asymmetry gets
produced only in the symmetric phase.
Expressing $n_L({\bf z})$ in terms of the Higgs number density
\begin{equation}
n_L=\frac{9k_q k_{\cal T}-8k_b k_{\cal T}
-5 k_b k_q}{k_{\cal H}(k_b+9 k_q+9 k_{\cal T})}\:{\cal H}
\end{equation}
and making use of Eqs.~(\ref{higgs1})-(\ref{bau}), we find that
\begin{equation}
\frac{n_B}{s}=-g(k_i)\frac{{\cal A}\overline{D}\Gamma_{{\rm ws}}}
{v_{\omega}^2 s},
\end{equation}
where $s=2\pi^2 g_{*s}T^3/45$ is the entropy density ($g_{*s}$
being
the effective number of relativistic degrees of freedom) and
$g(k_i)$
is a numerical coefficient depending upon the light degrees of
freedom present in the thermal bath.

We see that the final baryon asymmetry depends sensitively on the parameter ${\cal A}$, that is on the integral of the source. In  previous analysis \cite{nelson,noi}
the CP-violating sources have been  inserted into the diffusion equations by hand only after the CP-violating {\it currents} have been defined and computed. 
More specifically, CP-violating sources ${\cal S}$ associated to a generic charge density $j^0$ have been constructed from 
the current $j^\mu$  by the definition ${\cal S}=\partial_0 j^0$ \cite{nelson,noi}. 
This procedure has introduced an unacceptable   degree of  arbitrariness --different definitions of   CP-violating currents $j^\mu$ have been adopted  in the literature \cite{nelson,noi} and have lead to   different numerical results for the final baryon asymmetry. This is because the corresponding sources differ   in terms of   number of derivatives of the Higgs bubble wall profile and, therefore, in terms of  different powers of the bubble wall velocity $v_\omega$ and bubble wall width $L_\omega$.  

The  basic merit of the  CTP formalism is to provide us with a rigorous and self-consistent definition of the CP-violating sources  {\it within}  the quantum Boltzmann equations \cite{riotto97}.   The computation of the CP-violating {\it currents} for the right-handed stop and higgsino local densities  was  performed in \cite{noi} by means of the CTP formalism. Since   currents  were proportional to $\dot{\beta}$ in the thick bubble wall limit, {\it i.e.}  proportional to the first time derivative of the the Higgs profile,  sources turned out to be   proportional to the second time derivative of the Higgs profile \cite{noi}. However, the   derivation   of the transport QBE's   allows us to define the CP-violating sources uniquely, since they are automatically   incorporated  in the right-handed side of the QBE's \cite{riotto97}. 
This self-consistent procedure indicate that the sources in the quantum diffusion equations are proportional to the first time derivative of the Higgs configuration.   A comparison between the sources ${\cal S}_{\widetilde{t}_R}$, see Eq.   (\ref{source1}),  and ${\cal S}_{\widetilde{H}}$, see Eq.  (\ref{source2}),  obtained in the present work and the 
currents $j^0$ given  in  ref. \cite{noi} indicate that  they may be related
as
\begin{equation}
\label{z}
{\cal S}(T)\sim \frac{j^0(T)}{\tau}
\end{equation}
and it may be interpreted as the time derivative of the current density accumulated at the time
$T$ after  the wall has deposited at a given specific point the current density $j^0$ each interval $\tau$
\begin{equation}
{\cal S}(T)\sim \partial_0 \int^T dt \:\frac{j^0(t)}{\tau}.
\end{equation}
Here 
 $\tau=\Gamma^{-1}$ is the thermalization time of the right-handed stops and higgsinos, respectively. The integral over time is peculiar of the quantum approach and it induces memory effects. 

The parameter ${\cal A}$ computed from the sources (\ref{source1}) and (\ref{source2}) and  rigorously derived from the QBE's turns out to be\footnote{For large values of the left-handed stop mass 
the stop contribution to the baryon asymmetry is strongly
suppressed compared to the chargino and neutralino ones and 
the  numerical values of the baryon asymmetry depend linearly
on the phase of the Higgsino mass parameter.}
\begin{eqnarray}
{\cal A}&=&\frac{2 f(k_i)\Gamma_{\widetilde{H}}}{\overline{D} \; \lambda_{+}} \int_0^{\infty} du\; j_{\widetilde{H}}^0(u)
e^{-\lambda_+ u}\propto \frac{2 f(k_i)\Gamma_{\widetilde{H}}}{\overline{D} \; \lambda_{+}} I,\nonumber\\
I &\equiv &\int_0^{\infty} du\; v^2(u)\frac{d\beta(u)} {d u} e^{-\lambda_+ u}\simeq \int_0^{\infty} du\; v^2(u)\frac{d\beta(u)}{d u},
\end{eqnarray}
where $j_{\widetilde{H}}^0$ is the Higgsino current computed in \cite{noi},  
$\Gamma_{\widetilde{H}}$ is the higgsino damping rate and $f(k_i)$ is a coefficient depending upon the number of degrees of freedom present in the thermal bath. 

A comparison between this parameter ${\cal A}$ and the one obtained  in \cite{noi} indicates that memory effects introduce an enhancement factor in the final baryon asymmetry 
\begin{equation}
\label{new}
\frac{2\:\Gamma}{\lambda_{+}\:v_\omega}\simeq 70\:\left(\frac{\Gamma_{\widetilde{H}}}{5\times 10^{-2}\:T}\right)\left(\frac{0.1}{v_\omega}\right)
\end{equation}
with respect to the results given in \cite{noi} \footnote{The integral $I$ has been computed  including two-loop effects in ref. \cite{mariano} and results to be $I\simeq 10^{-2}$ for  $m_A = 150$--200 GeV. The same value (in absolute value) of the integral was obtained  in ref. \cite{noi} by making use of a reasonable Ansatz for the Higgs profile. The reason is that this result was based on
the temperature $T_0$ defined to be the temperature at which the Higgs 
potential becomes flat at the origin. This temperature  is lower than the temperature $T_c$ at which the tunneling process occurs and the numerical result of ref. \cite{mariano} indicates that 
 the integral $I$  is a  decreasing function of the temperature.
Since the   results of ref. \cite{noi} were based on the one-loop effective potential, the  result in \cite{mariano}  is  accidentally very close to the numerical result of \cite{noi}.
Therefore, the numerical difference between the value of the final baryon asymmetry obtained here and the one evaluated in ref. \cite{noi} reduces to the 
expression (\ref{new}). We thank M. Quiros for discussions on this point.
}.  
Our results indicate that memory effects play a crucial role  in relaxing the stringent lower bound on the values of the CP-violating phases obtained in previous analysis \cite{noi,ck}. In particular, the observed baryon asymmetry $n_B/s\simeq 4\times 10^{-11}$ may be originated by Higgsinos if the CP-violating phase $\phi_\mu$  of the $\mu$-parameter is 
\begin{equation}
|\sin(\phi_\mu)|\gsim 10^{-3}\left(\frac{v_\omega}{0.1}\right).
\end{equation}
This lower bound  is two orders of magnitude weaker than the one obtained in previous works \cite{noi,ck}.  
These small values of the phase $\phi_\mu$ are consistent with the constraints from the electric dipole moment of the neutron even if the squarks of the first and second generation have masses of the order of 100 GeV. In the limit of thick bubble walls,   the CP-violating sources are also characterized by resonance effects \cite{noi,riotto97} when the particles involved in the construction of the source are degenerate in mass.  The resonance is manifest in the function $G$ defined in (\ref{v1}). The interpretation of the resonance is rather straightforward if we think  in terms of  scatterings of the quasiparticles off the advancing low momentum bubble wall configuration.  For momenta of order of the critical
temperature, the scattering is more efficient   when the Higgsinos and 
gauginos are nearly degenerate in mass, $\mu\sim M_2$.   
A similar effect has been found in ref. \cite{ck} where the system was  studied  in the classical limit. These classical treatments somehow obscure the origin of the CP-violating effects as resulting from quantum interference and the origin of the resonance.  Moreover, memory effects are not  present in the classical approach which, therefore, understimates the value of the final baryon asymmetry.

\newpage

\begin{flushleft}
{\bf  8. Conclusions}
\end{flushleft}

In conclusion, we have reanalyzed  the computation
 of
the baryon asymmetry generated at the electroweak phase transition
in the Minimal Supersymmetric Standard Model. We have argued that the   CTP formalism is the natural  guide towards a rigorous and self-consistent definition of the CP-violating sources   within  the quantum Boltzmann equations.  The QBE's manifest --in the scattering terms --non-Markovian features. In the limit of thick bubble wall, however, a number of approximations may be 
performed so that the QBE's get reduced to the more familiar diffusion/Boltzmann equations.  Strong memory effects  enhance the strength of the CP-violating sources by at least two orders of magnitude. The 
baryon asymmetry is mainly generated by  Higgsinos, provided that Higgsinos and gauginos are not much
heavier than the electroweak critical temperature and the phase $\phi_\mu$ is larger than $10^{-3}$. This value is further reduced for $v_\omega\lsim 0.1$. 
It is intriguing that these small  values of the phases
are perfectly  consistent with the constraints from the
electric dipole moment of the neutron and   squarks of the
first and second generation as light as $\sim 100$ GeV may be tolerated.

\vskip1cm
{\large\bf Acknowledgements}
\vskip 0.2cm
The author would like to thank M. Carena,  M. Quiros and C.E.M. Wagner for useful discussions.

\def\NPB#1#2#3{Nucl. Phys. {\bf B#1}, #3 (19#2)}
\def\PLB#1#2#3{Phys. Lett. {\bf B#1}, #3 (19#2) }
\def\PLBold#1#2#3{Phys. Lett. {\bf#1B} (19#2) #3}
\def\PRD#1#2#3{Phys. Rev. {\bf D#1}, #3 (19#2) }
\def\PRL#1#2#3{Phys. Rev. Lett. {\bf#1} (19#2) #3}
\def\PRT#1#2#3{Phys. Rep. {\bf#1} (19#2) #3}
\def\ARAA#1#2#3{Ann. Rev. Astron. Astrophys. {\bf#1} (19#2) #3}
\def\ARNP#1#2#3{Ann. Rev. Nucl. Part. Sci. {\bf#1} (19#2) #3}
\def\MPL#1#2#3{Mod. Phys. Lett. {\bf #1} (19#2) #3}
\def\ZPC#1#2#3{Zeit. f\"ur Physik {\bf C#1} (19#2) #3}
\def\APJ#1#2#3{Ap. J. {\bf #1} (19#2) #3}
\def\AP#1#2#3{{Ann. Phys. } {\bf #1} (19#2) #3}
\def\RMP#1#2#3{{Rev. Mod. Phys. } {\bf #1} (19#2) #3}
\def\CMP#1#2#3{{Comm. Math. Phys. } {\bf #1} (19#2) #3}

% BODY

\end{document}